\documentstyle[11pt]{article}
\def\one{1\hskip -.37em 1}     
\def\proj{E\hskip -.67em I}     
\begin{document}
\begin{titlepage}
\begin{centering}
 
{\ }\vspace{2cm}
 
{\Large\bf The Physical Projector and\\}
\vspace{0.5cm}
{\Large\bf Topological Quantum Field Theories:}\\
\vspace{0.5cm}
{\Large\bf U(1) Chern-Simons Theory in 2+1 Dimensions}\\
\vspace{2cm}
Jan Govaerts\footnote{E-mail: {\tt govaerts@fynu.ucl.ac.be}}
and Bernadette Deschepper\\
\vspace{1.0cm}
{\em Institut de Physique Nucl\'eaire}\\
{\em Universit\'e catholique de Louvain}\\
{\em 2, Chemin du Cyclotron}\\
{\em B-1348 Louvain-la-Neuve, Belgium}\\
\vspace{2cm}
\begin{abstract}

\noindent The recently proposed physical projector approach to the quantisation
of gauge invariant systems
is applied to the U(1) Chern-Simons theory in 2+1 dimensions as
one of the simplest examples of a topological quantum field theory. 
The physical projector is explicitely demonstrated to be capable of effecting 
the required projection from the initially infinite number of degrees of
freedom to the finite set of gauge invariant physical states whose properties
are determined by the topology of the underlying manifold.

\end{abstract}

\vspace{15pt}


\end{centering} 

\vspace{85pt}

\noindent PACS numbers: 11.15.-q

\vspace{20pt}

\noindent hep-th/9909221\\
\noindent September 1999

\end{titlepage}

\section{Introduction}
\label{Sect1}

The general gauge invariance principle pervades all of modern physics
at the turn of this century, as a most basic conceptual principle unifying
the fields of algebra, topology and geometry with those of the fundamental
interactions of the elementary quantum excitations in the natural Universe.
This fascinating convergence of ideas is probably nowhere else better
demonstrated than within the recent developments of M-theory as the prime
(and sole) candidate for a fundamental unification\cite{Schwarz}.
Given the many mathematics and physics
riches hidden deep into the structures and dynamics of gauge
invariant theories, a thorough understanding sets a genuine challenge to
the methods developed over the years in order to address such 
issues. For example, a manifest realisation of the gauge invariance principle
requires the presence among the degrees of freedom of such systems of 
redundant variables whose dynamics is specified
through arbitrary functions characterizing the gauge freedom inherent to
the description. This situation leads to specific problems, especially
when quantising such theories, since some gauge fixing procedure has to be
applied in order to effectively remove in a consistent way
the contributions of gauge variant states to physical observables. 
Most often than not, such gauge fixings suffer
Gribov problems\cite{Gribov,Singer,Gov1}, which must properly 
be addressed if one is to correctly
account for the quantum dynamics of gauge invariant systems, certainly
within a non perturbative framework.
Among gauge theories, topological quantum field theories\cite{Witten1,Witten2}
provide the most extreme example of such a situation,
since their infinite number of degrees of freedom includes only a
{\em finite\/} number of gauge in\-va\-riant physical states, whose
properties are in addition solely determined by the topology of 
the underlying manifold irrespective of its geometry.

In a recent development\cite{Klauder1}, 
a new approach to the quantisation of gauge theories
was proposed, which avoids from the outset any gauge fixing and thus any
issue of the eventuality of some Gribov problem\cite{Gov2}. This approach 
is directly set within the necessary framework of Dirac's quantisation of 
constrained systems\cite{Gov1}. No extension---with its cort\`ege of 
ghosts and ghosts for ghosts---or reduction of the original set of degrees of
freedom is required, as is the case for all approaches which necessarily
implement some gauge fixing procedure with its inherent risks of
Gribov problems\cite{Gov1}. Nonetheless, the correct representation of the true
quantum dynamics of the system is achieved in this new approach, which uses
in an essential way the projection operator\cite{Klauder1} onto the 
subspace of gauge invariant physical states of a given gauge invariant system.
Some of the advantages of the physical projector approach have already
been explored and demonstrated in a few simple quantum mechanical gauge
invariant systems\cite{Klauder2,Villa}. In the present work, we wish 
to illustrate how the same methods are capable to also deal with the 
intricacies of topological quantum field theories which, even though 
possessing only a finite set of physical states, require an infinite number 
of degrees of freedom and states for their formulation. Indeed, it will be 
shown that the physical projector precisely effects this required projection.

The specific case addressed here as one of the simplest possibilities, is that
of the pure Chern-Simons theory in 2+1 dimensions with gauge group U(1).
Moreover, the discussion will be made explicit when the topology of the
underlying manifold is that of $\Sigma\times I\!\!R$, where $\Sigma$ is
a two dimensional compact Riemann surface taken to be
a 2-torus $T_2$ for most of our considerations. This system has been studied
from quite a many different points of view\cite{Witten2,Bos1,Bos2,Lab,CS,Dunne}. 
The consistency of the physical
projector approach will be demonstrated by deriving again some of the
same results through an explicit resolution of the gauge invariant 
quantum dynamics within this specific framework which avoids any gauge 
fixing whatsoever and thus also any Gribov problem.

The outline of the discussion is as follows. Sect.\ref{Sect2} briefly elaborates
on the classical constrained Hamiltonian formulation for Chern-Simons 
theories with arbitrary gauge symmetry group. These considerations are then 
particularised in Sect.\ref{Sect3} to the U(1) case restricted to the 
$I\!\!R\times T_2$ topology, enabling a straighforward Fourier mode analysis 
of the then discrete infinite set of degrees of freedom. In Sect.\ref{Sect4}, 
the Dirac quantisation of the system is developed, leading in Sect.\ref{Sect5}
to the construction of the physical projector. These results are then
explicitely used in Sect.\ref{Sect6} in order to identify the spectrum 
of physical states in the U(1) theory and to determine their
coherent state wave function representations. Finally, the discussion
ends with Conclusions, while some necessary details are included in 
an Appendix in order not to detract from the main line of arguments.

\section{Classical Chern-Simons Theories}
\label{Sect2}

Let $G$ be a compact simple Lie group of hermitian generators $T^a$ 
($a=1,2,\cdots,{\rm dim}\ G$) and structure constants $f^{abc}$ such that
$[T^a,T^b]=if^{abc}T^c$. In terms of the gauge connection $A^a_\mu$,
the action for the associated 2+1 dimensional Chern-Simons theory is 
then given by
\begin{equation}
\begin{array}{l c r}
S&=&N_k\int_{I\!\!R\times\Sigma}dx^0dx^1dx^2\,
\epsilon^{\mu\nu\rho}\Big[A^a_\mu\partial_\nu A^a_\rho-\frac{1}{3}f^{abc}
A^a_\mu A^b_\nu A^c_\rho\Big]\\
 & & \\
&=&\frac{1}{2}N_k\int_{I\!\!R\times\Sigma}dx^0dx^1dx^2\,
\epsilon^{\mu\nu\rho}\Big[A^a_\mu F^a_{\nu\rho}+\frac{1}{3}f^{abc}
A^a_\mu A^b_\nu A^c_\rho\Big]\ \ ,
\end{array}
\label{eq:S1}
\end{equation}
where $F^a_{\mu\nu}=\partial_\mu A^a_\nu-\partial_\nu A^a_\mu-
f^{abc}A^b_\mu A^c_\nu$, $\epsilon^{012}=+1$, while
$N_k$ is a normalisation factor (the usual gauge coupling constant $g$ has
been absorbed into the gauge connection $A^a_\mu$).

As is well known, under small gauge transformations ({\em i.e.\/} 
continuously connected to the identity transformation), the Lagrangian density
in (\ref{eq:S1}) remains invariant up to a surface term. For large gauge
transformations however ({\em i.e.\/} those transformations in a homotopy
class different from that of the identity transformation), the action
(\ref{eq:S1}) changes by a term proportional to a topological invariant, 
namely the winding number of the gauge transformation. Consequently, at the 
quantum level, invariance under large gauge transformations
requires that the normalisation factor $N_k$ be quantised,
which is the reason for its notation. Note also that 
no metric structure whatsoever is necessary for the definition of the action 
(\ref{eq:S1}). Indices $\mu,\nu,\rho=0,1,2$ are not raised nor lowered, 
while the notation, though reminiscent of a Minkowski signature metric, 
is in fact related to the specific topology $I\!\!R\times\Sigma$ considered 
for the three dimensional manifold. The choice of the real line $I\!\!R$ 
for the time evolution coordinate $x^0$ is made for the purpose of canonical 
quantisation hereafter, while
any other type of compact topology for the three dimensional manifold
may be obtained from $I\!\!R\times\Sigma$ through gluing and 
twisting\cite{Witten2}.

Hence, (\ref{eq:S1}) defines a topological field theory, namely a field theory
whose gauge freedom is so large that its gauge invariant configurations
characterize solely the topology of the underlying manifold, irrespective
of its geometry\cite{Witten1}. In the present case, this is made obvious 
in terms of the associated equations of motion,
\begin{equation}
\epsilon^{\mu\nu\rho}F^a_{\nu\rho}=0\ \ \ \Leftrightarrow\ \ \
F^a_{\mu\nu}=0\ \ \ .
\end{equation}
Indeed, the modular space of flat gauge connections on the base manifold
is finite dimensional and purely topological in its characterization through
holonomies around the non-contractible cycles in the manifold. 
Quantisation of the Chern-Simons theory thus defines
the quantisation of a system whose configuration space---which actually 
coincides with its phase space---is this modular space of flat connections.

The action (\ref{eq:S1}) being of first-order form in time derivatives
of fields, is already in the Hamiltonian form necessary for canonical
quantisation\cite{Faddeev,Gov1}. Indeed, we explicitely have,
\begin{equation}
S=\int d^3x^\mu\Big[\partial_0A^a_i\,N_k\epsilon^{ij}A^a_j+
A^a_0\,N_k\epsilon^{ij}F^a_{ij}-
\partial_i\left(N_k \epsilon^{ij}A^a_jA^a_0\right)\Big]\ \ \ ,
\label{eq:S2}
\end{equation}
where $\epsilon^{ij}$ $(i,j=1,2)$ is the two-dimensional antisymmetric
symbol with $\epsilon^{12}=+1$. Consequently, the actual phase space of the
system consists of the field components $A^a_i$ $(i=1,2)$ which form a
pair of conjugate variables with symplectic structure defined by the brackets
\begin{equation}
\{A^a_1(\vec{x},x^0),A^b_2(\vec{y},x^0)\}=\frac{1}{2N_k}\,\delta^{ab}\,
\delta^{(2)}(\vec{x}-\vec{y})\ \ \ .
\label{eq:brackets}
\end{equation}
In addition, the first-class Hamiltonian of the system vanishes identically,
$H=0$, as befits any system invariant under local coordinate
reparametrisations, while finally the time components $A^a_0$ of the gauge 
connection are the Lagrange multipliers for the first-class constraints
\begin{equation}
\phi^a=-2N_kF^a_{12}=-2N_k\left[\partial_1A^a_2-
\partial_2A^a_1-f^{abc}A^b_1A^c_2\right]\ \ \ ,
\end{equation}
whose algebra of brackets is that of the gauge group $G$,
\begin{equation}
\{\phi^a(\vec{x},x^0),\phi^b(\vec{y},x^0)\}=f^{abc}\phi^c(\vec{x},x^0)
\delta^{(2)}(\vec{x}-\vec{y})\ \ \ .
\end{equation}
That these constraints are indeed the local generators of small gauge 
transformations is confirmed through their infinitesimal action on the phase 
space variables $A^a_i$,
\begin{equation}
\delta_\theta A^a_i(\vec{x},x^0)=\Big\{A^a_i(\vec{x},x^0)\ ,\
\int_\Sigma d^2\vec{y}\,\theta^b(\vec{y},x^0)\phi^b(\vec{y},x^0)\Big\}=
\partial_i\theta^a(\vec{x},x^0)+
f^{abc}\theta^b(\vec{x},x^0)A^c_i(\vec{x},x^0)\ \ \ ,\ \ \
\end{equation}
while the Lagrange multipliers $A^a_0$ must vary according to
\begin{equation}
\delta_\theta A^a_0(\vec{x},x^0)=
\partial_0\theta^a(\vec{x},x^0)+f^{abc}\theta^b(\vec{x},x^0)A^c_0(\vec{x},x^0)
\ \ .
\end{equation}
The constraints $\phi^a$ also coincide, up to surface terms, with the
Noether charge densities related to the gauge symmetry. Indeed, 
the Noether currents
$\gamma^{a\mu}=\epsilon^{\mu\nu\rho}N_kf^{abc}A^b_\nu A^c_\rho$
are conserved for solutions to the equations of motion,
$\partial_\mu\,\gamma^{a\mu}=0$, so that the associated charges read
\begin{equation}
Q^a=\int_\Sigma d^2\vec{x}\,\gamma^{a(\mu=0)}=
\int_\Sigma d^2\vec{x}\,
\left[\phi^a+2N_k(\partial_1A^a_2-\partial_2A^a_1)\right]\ \ \ .
\label{eq:Qa}
\end{equation}

The above conclusions based on the first-order action (\ref{eq:S2}) are of
course confirmed through an explicit application of Dirac's algorithm for
the construction of the Hamiltonian formulation of constrained 
systems\cite{Gov1}.
In particular, the brackets (\ref{eq:brackets}) then correspond to
Dirac brackets after second-class constrainst have been solved for, while
(\ref{eq:S2}) then describes the so-called ``fundamental Hamiltonian
formulation"\cite{Gov1} of any dynamical system. Incidentally, note that 
surface terms which appear in (\ref{eq:S2}) and (\ref{eq:Qa})
are irrelevant for such a Hamiltonian construction, which is in essence
a local construct on the manifold $\Sigma$. In any case, they do not
contribute when $\Sigma$ is without boundaries.

\section{The U(1) Theory on the Torus} 
\label{Sect3}

Henceforth, we shall restrict
the discussion to the gauge group $G=$U(1) and to the compact Riemann manifold
$\Sigma$ being the two dimensional torus $T_2$. This choice is made for the
specific purpose of demonstrating that the physical projector approach
is capable of properly quantising such theories in the simplest of cases,
leaving more general choices to be explored elsewhere with the same
techniques. In particular, the torus mode expansions to be specified
presently may be extended to Riemann surfaces $\Sigma$ of arbitrary genus
through the use of abelian differentials and the Krichever-Novikov
operator formalism\cite{Bos1,Bos2,Lab,Krichever,Vafa,Bonora}. 
The extension to non abelian gauge groups $G$ requires
further techniques of coherent states not included in the present discussion.

Given the manifold $\Sigma=T_2$, let us consider the local trivialisation
of this topology as\-so\-cia\-ted to a choice of basis of its first homology
group with cycles $a$ and $b$. Correspondingly, the choice of
local coordinates $x^1$ and $x^2$ is such that $0 < x^1, x^2 < 1$.
Related to this trivialisation of $T_2$, fields over $T_2$ may be Fourier 
expanded, so that the total number of degrees of freedom, though infinite, is
represented in terms of a {\em discrete\/} set of modes over $T_2$. Explicitely,
in a real parametrisation we have (from here on, any dependency on $x^0$
is left implicit while the single index $a=1$ for $G=$U(1) is not displayed),
\begin{equation}
\begin{array}{r c l}
A_i(\vec{x})&=&\sum_{n_1=0}^{+\infty}\sum_{n_2=0}^{+\infty}
A_i^{++}(n_1,n_2)\cos 2\pi n_1x^1\,\cos 2\pi n_2x^2\ +\ \\
 & & \\
&+&\sum_{n_1=0}^{+\infty}\sum_{n_2=1}^{+\infty}
A_i^{+-}(n_1,n_2)\cos 2\pi n_1x^1\,\sin 2\pi n_2x^2\ +\ \\
 & & \\
&+&\sum_{n_1=1}^{+\infty}\sum_{n_2=0}^{+\infty}
A_i^{-+}(n_1,n_2)\sin 2\pi n_1x^1\,\cos 2\pi n_2x^2\ +\ \\
 & & \\
&+&\sum_{n_1=1}^{+\infty}\sum_{n_2=1}^{+\infty}
A_i^{--}(n_1,n_2)\sin 2\pi n_1x^1\,\sin 2\pi n_2x^2\ \ \ .
\end{array}
\label{eq:Amode}
\end{equation}
Let us emphasize that such expansions do not include the terms which 
would be associated to the following modes: $A_i^{+-}(n_1,0)$, 
$A_i^{-+}(0,n_2)$, $A_i^{--}(n_1,0)$ and $A_i^{--}(0,n_2)$ 
($n_1,n_2=1,2,\dots$). Similar mode expansions---to which the remark 
just made also applies---are 
obtained for any quantity defined over $T_2$. For the U(1) generator 
$\phi(\vec{x})$, one has (for a non-abelian group $G$, terms bilinear in 
the $A_i^{\pm\pm}(n_1,n_2)$ modes also contribute, which is the reason for our 
restriction to $G=$U(1))
\begin{equation}
\begin{array}{r c l}
\phi^{++}(n_1,n_2)&=&-4\pi N_k\Big[+n_1A_2^{-+}(n_1,n_2)-n_2A_1^{+-}(n_1,n_2)
\Big]\ \ ,\\
 & & \\
\phi^{+-}(n_1,n_2)&=&-4\pi N_k\Big[+n_1A_2^{--}(n_1,n_2)+n_2A_1^{++}(n_1,n_2)
\Big]\ \ ,\\
 & & \\
\phi^{-+}(n_1,n_2)&=&-4\pi N_k\Big[-n_1A_2^{++}(n_1,n_2)-n_2A_1^{--}(n_1,n_2)
\Big]\ \ ,\\
 & & \\
\phi^{--}(n_1,n_2)&=&-4\pi N_k\Big[-n_1A_2^{+-}(n_1,n_2)+n_2A_1^{-+}(n_1,n_2)
\Big]\ \ .
\end{array}
\end{equation}

Since $\{\phi(\vec{x}),\phi(\vec{y})\}=0$ in the abelian U(1) case, 
all the modes $\phi^{\pm\pm}(n_1,n_2)$ have vanishing brackets with one
another, while those for the phase space modes $A_i^{\pm\pm}(n_1,n_2)$ are 
given by
\begin{equation}
\{A_1^{\pm\pm}(n_1,n_2),A_2^{\pm\pm}(m_1,m_2)\}=\frac{2}{N_k}\,
f^{\pm\pm}(n_1,n_2)\,\delta_{n_1,m_1}\,\delta_{n_2,m_2}\ \ \ ,
\label{eq:modebrackets}
\end{equation}
where
\begin{equation}
\begin{array}{r c l c r c l}
f^{++}(n_1,n_2)&=&\frac{1}{(1+\delta_{n_1,0})(1+\delta_{n_2,0})}\ \ &,&\ \
f^{+-}(n_1,n_2)&=&\frac{1-\delta_{n_2,0}}{1+\delta_{n_1,0}}\ \ ,\ \ \\
 & & & & & \\
f^{-+}(n_1,n_2)&=&\frac{1-\delta_{n_1,0}}{1+\delta_{n_2,0}}\ \ &,&\ \
f^{--}(n_1,n_2)&=&(1-\delta_{n_1,0})(1-\delta_{n_2,0})\ .
\end{array}
\label{eq:fn}
\end{equation}

In order to understand how small and large gauge transformations---the latter
not being generated by the first-class constraint $\phi(\vec{x})$---are
represented in terms of these mode expansions, let us
consider the general gauge transformation of the field
$A_\mu$, namely $A'_\mu=A_\mu+\partial_\mu\theta$, as\-so\-cia\-ted to the U(1) 
local phase transformation $U(\vec{x},x^0)=e^{i\theta(\vec{x},x^0)}$. 
A point central to our discussion is that the arbitrary function 
$\theta(\vec{x},x^0)$ may always be expressed as
\begin{equation}
\theta(\vec{x},x^0)=\theta_0(\vec{x},x^0)+2\pi k_1x^1+2\pi k_2x^2\ \ \ ,
\label{eq:theta1}
\end{equation}
where $\theta_0(\vec{x},x^0)$ is an arbitrary {\em periodic\/} 
function---{\em i.e.\/} a {\em scalar field\/} on $T_2$---while $k_1$
and $k_2$ are arbitrary positive or negative integers. Indeed, any
small gauge transformation is defined in terms of some function 
$\theta_0(\vec{x},x^0)$ with $(k_1,k_2)=(0,0)$, while any large gauge 
transformation $\theta(\vec{x},x^0)$ may always
be brought to the above general form with some specific function 
$\theta_0(\vec{x},x^0)$, the integers $k_1$ and $k_2$ then labelling
the U(1) holonomies of the gauge transformation around the chosen $a$ and $b$
homology cycles in $T_2$. Any gauge transformation thus falls into such
a $(k_1,k_2)$ homotopy class of the gauge group over $T_2$.
In terms of the mode expansion of the gauge parameter function 
(\ref{eq:theta1}), the non-zero modes $A_i^{\pm\pm}(n_1,n_2)$ 
($n_1\ne 0$ or $n_2\ne 0$, or both) are transformed according to
\begin{equation}
\begin{array}{r c l c r c l}
\Delta A_1^{++}(n_1,n_2)&=&+2\pi n_1\theta_0^{-+}(n_1,n_2)
\ \ &,&\ \ 
\Delta A_2^{++}(n_1,n_2)&=&+2\pi n_2\theta_0^{+-}(n_1,n_2)
\ \ ,\\
 & & & & & & \\
\Delta A_1^{+-}(n_1,n_2)&=&+2\pi n_1\theta_0^{--}(n_1,n_2)
\ \ &,&\ \ 
\Delta A_2^{+-}(n_1,n_2)&=&-2\pi n_2\theta_0^{++}(n_1,n_2)
\ \ ,\\
 & & & & & & \\
\Delta A_1^{-+}(n_1,n_2)&=&-2\pi n_1\theta_0^{++}(n_1,n_2)
\ \ &,&\ \ 
\Delta A_2^{-+}(n_1,n_2)&=&+2\pi n_2\theta_0^{--}(n_1,n_2)
\ \ ,\\
 & & & & & & \\
\Delta A_1^{--}(n_1,n_2)&=&-2\pi n_1\theta_0^{+-}(n_1,n_2)
\ \ &,&\ \ 
\Delta A_2^{--}(n_1,n_2)&=&-2\pi n_2\theta_0^{-+}(n_1,n_2)
\ \ ,\\
\end{array}
\label{eq:gauge1}
\end{equation}
where
$\Delta A_i^{\pm\pm}(n_1,n_2)=\left(A_i^{\pm\pm}(n_1,n_2)\right)^\prime-
A_i^{\pm\pm}(n_1,n_2)$, while the zero modes $A_i^{++}(0,0)$ transform as
\begin{equation}
\Delta A_1^{++}(0,0)=2\pi k_1\ \ \ ,\ \ \
\Delta A_2^{++}(0,0)=2\pi k_2\ \ \ .
\label{eq:gauge2}
\end{equation}

These different expressions thus nicely establish that small gauge
transformations---ge\-ne\-ra\-ted by the first-class constraint 
$\phi(\vec{x})$---modify only the non-zero modes and that large gauge
transformations only affect the zero modes of the gauge connection 
$A_i(\vec{x})$. Turning the argument around, one thus concludes that 
the system factorizes into two types of degrees of freedom, namely the 
non-zero modes $A^{\pm\pm}_i(n_1,n_2)$ ($n_1\ne 0$ or $n_2\ne 0$) directly 
related to small gauge transformations only, and the zero modes $A^{++}_i(0,0)$
directly related to large gauge transformations only. Moreover, the above
expressions also show that it is always possible to set half the non-zero
modes to zero by an appropriate small gauge transformation, namely either
the $i=1$ or the $i=2$ component for each of the modes $A^{\pm\pm}_i(n_1,n_2)$
(the choice of which of these two components is set to zero is left open for
the modes with both $n_1\ne 0$ and $n_2\ne 0$, but not for those modes
for which either $n_1=0$ or $n_2=0$). Consequently, invariance under small
gauge transformations implies that the physical content of the system
actually reduces to that of its zero mode sector $A^{++}_i(0,0)$ ($i=1,2$)
only, while the physical content of its non-zero mode sector is gauge
equi\-va\-lent to the trivial
solution $A_i(\vec{x})=0$ to the flat connection condition $F_{12}(\vec{x})=0$ 
associated to the vanishing holonomies $(k_1,k_2)=(0,0)$. That the 
physics of these systems lies entirely in their zero mode sector remains 
valid at the quantum level as well, as is shown hereafter
(in fact, this conclusion also extends to Riemann surfaces $\Sigma$ of 
arbitrary genus and for any choice of non-abelian gauge group 
$G$\cite{Witten2,Bos1,Bos2,Lab,CS}).

An identical separation also applies to the modes of the gauge parameter 
function $\theta(\vec{x},x^0)$. As shown above, the term 
$(2\pi k_1x^1+2\pi k_2x^2)$ corresponds to large gauge transformations only, 
while the contribution $\theta_0(\vec{x},x^0)$ induces small transformations 
only, whose zero mode $\theta^{++}_0(0,0)$ in fact completely decouples. Indeed,
the latter mode corresponds to a global phase transformation, which for the
real degrees of freedom $A_\mu(\vec{x},x^0)$ stands for no transformation
at all. In other words, in as far as the small gauge parameter function
$\theta_0(\vec{x},x^0)$ is concerned, one could say that its zero mode
$\theta^{++}_0(0,0)$ has in fact been traded for the $(k_1,k_2)$
parameters characterizing the holonomies of a large gauge transformation.
The zero mode $\theta_0^{++}(0,0)$ for small gauge parameter functions
$\theta_0(\vec{x})$ thus does not enter our considerations
in any way whatsoever, and may always be set to zero.
Finally, let us simply point out also that the range of each of the non-zero
modes $\theta^{\pm\pm}_0(n_1,n_2)$ is the entire real line, running from
$-\infty$ to $+\infty$ (as opposed to the zero mode $\theta^{++}_0(0,0)$ 
which would have taken its values in the interval $[0,2\pi]$ for example,
had it contributed to gauge symmetries of the system).

It may appear that by using the mode expansion (\ref{eq:Amode}), the
gauge field $A_i(\vec{x})$ is assumed to obey periodic boundary conditions,
which would amount to implicitely assuming that $A_i(\vec{x})$ defines a vector
field over $T_2$ rather than a possibly non trivial section of a U(1) bundle 
over $T_2$, the latter case being associated to the possibility of twisted 
boundary conditions for the components $A_i(\vec{x})$\cite{tHooft}. However, 
the local trivialisation of $T_2$ in terms of the coordinates $0<x^1,x^2<1$
does not define a complete covering of $T_2$, which requires at least
$2^2=4$ different overlapping coordinate charts. Moving from one chart to
another implies that the gauge connection $A_i(\vec{x})$ changes also by
gauge transformations which include large ones associated to
twisted boundary conditions. Therefore, by explicitely considering in our
definition of the gauge invariant system large gauge transformations,
the possibility of twisted boundary conditions is implicitely included,
and the above Fourier mode decomposition of the phase space degrees of freedom
$A_i(\vec{x})$ is fully warranted.

\section{Dirac Quantisation} 
\label{Sect4}

Given the above mode decompositions of the
Hamiltonian formulation of the system defined over the torus $T_2$, 
its canonical quantisation proceeds
straightforwardly through the correspondence principle according to
which brackets now correspond to commutation relations being set equal to the 
value of the bracket multiplied by $i\hbar$. Thus, given 
(\ref{eq:modebrackets}), the
fundamental quantum operators are the modes $\hat{A}_i^{\pm\pm}(n_1,n_2)$
such that
\begin{equation}
[\hat{A}_1^{\pm\pm}(n_1,n_2),\hat{A}_2^{\pm\pm}(m_1,m_2)]=
\frac{2i\hbar}{N_k} f^{\pm\pm}(n_1,n_2)\delta_{n_1,m_1}\delta_{n_2,m_2}\ \ \ ,
\end{equation}
with in particular for the zero modes,
\begin{equation}
\Big[\hat{A}_1^{++}(0,0),\hat{A}_2^{++}(0,0)\Big]=i\frac{\hbar}{2N_k}\ \ \ .
\label{eq:commzero}
\end{equation}
Henceforth, we shall thus assume explicitely that $N_k>0$, with the
understanding that the case when $N_k<0$ is then obtained simply by 
interchanging the r\^oles of the coordinates $x^1$ and $x^2$.

It is already possible at this stage to determine the
number of quantum physical states\cite{Witten2}. Indeed, the relations
(\ref{eq:gauge1}) and (\ref{eq:gauge2}) show that the actual gauge invariant
phase space degrees of freedom are the zero modes $A_i^{++}(0,0)$
defined up to integer shifts by $2\pi$, while half the non-zero
modes---either the $i=1$ or $i=2$ component, depending on the given
mode---may be set to zero through small gauge transformations. In other
words, the actual physical phase space of the system is a two dimensional
torus of volume $(2\pi)^2$, an instance of a phase space which is not
a cotangent bundle as is usually the case but rather a compact manifold. 
The quantisation of the system thus amounts to quantising this two dimensional 
torus, with the commutation relation (\ref{eq:commzero}) in which 
$\hbar'=\hbar/(2N_k)$ plays
the r\^ole of an effective Planck constant. In particular, the total number
of physical states is thus given by the volume $(2\pi)^2$ of phase space
divided by that of each quantum cell $(2\pi\hbar')$ for the
degree of freedom $A_1^{++}(0,0)$, namely
\begin{equation}
\frac{(2\pi)^2}{2\pi(\hbar/2N_k)}=\frac{4\pi}{\hbar}N_k\ \ \ .
\end{equation}
Consequently, the normalisation factor $N_k$ ought to be quantised with a value 
$N_k=\hbar k/(4\pi)$ to be associated with $k$ physical states ($k=1,2,\dots$).
Precisely this quantisation condition is established hereafter
by considering large gauge transformations of the system; this quantisation
condition will then be specified further 
later on when considering modular transformations of the underlying
torus $T_2$, which then require the integer $k$ to also be even.

The above commutation relations for the modes $\hat{A}_i^{\pm\pm}(n_1,n_2)$
define an infinite tensor pro\-duct of Heisenberg algebras. In order to set up
a coherent state representation through creation and
annihilation operators associated to this Heisenberg algebra, it is
necessary to introduce a complex structure on the initial base manifold
$T_2$. The necessity of introducing some further structure on $\Sigma$ beyond 
the purely topological one is also unavoidable in all other quantisation 
frameworks for Chern-Simons theories\cite{Witten2}. In fact, all other 
approaches require a metric structure on $\Sigma$, while it is then shown that 
the quantised system nevertheless depends only on the complex structure 
(or conformal class of the metric) on $\Sigma$, with the space of gauge 
invariant physical states providing a projective representation of 
the modular group of $\Sigma$ due to a quantum conformal anomaly\cite{Witten2}. 
This is how close a topological quantum 
field theory may come to being purely topological. In the present approach,
the necessity of introducing a complex structure over $\Sigma$ is thus
seen to arise from a coherent state quantisation of the system, while,
in contradistinction to other quantisation methods,
it is also gratifying to realise that no further structure is required within 
this approach since the quantised system should in any case turn out
to be independent of any such additional structure.

On $T_2$, a complex structure is characterized through a complex
parameter $\tau=\tau_1+i\tau_2$ whose imaginary part is strictly positive,
($\tau_2>0$), with the modular group
$PSL(2,Z\!\!\!Z)$ of transformations generated by $(T: \tau\rightarrow\tau+1)$
and $(S: \tau\rightarrow -1/\tau)$ defining the classes of inequivalent
complex structures under global diffeomeorphisms in $T_2$. Associated to the
complex parametrisation,
\begin{equation}
z=x^1+\tau x^2\ \ \ ,\ \ \ 
dz\,d\bar{z}=|dx^1+\tau dx^2|^2=(dx^1)^2+2\tau_1 dx^1 dx^2
+|\tau|^2(dx^2)^2\ \ \ ,
\end{equation}
the gauge connection 1-form reads as
\begin{equation}
A=dx^1 A_1+dx^2 A_2=dz A_z + d\bar{z} A_{\bar{z}}\ \ \ ,
\end{equation}
with
\begin{equation}
A_z=\frac{i}{2\tau_2}\Big[\bar{\tau}A_1-A_2\Big]\ \ \ ,\ \ \ 
A_{\bar{z}}=-\frac{i}{2\tau_2}\Big[\tau A_1-A_2\Big]\ \ \ .
\label{eq:Az}
\end{equation}

Given the choice of complex structure parametrised by $\tau$, the
annihilation operators for the quantised system are defined by
\begin{equation}
\alpha^{\pm\pm}(n_1,n_2)=\sqrt{\frac{1}{f^{\pm\pm}(n_1,n_2)}
\frac{N_k}{4\hbar\tau_2}}\
\Big[-i\tau \hat{A}_1^{\pm\pm}(n_1,n_2)+i\hat{A}_2^{\pm\pm}(n_1,n_2)\Big]\ \ \ ,
\label{eq:annihilation}
\end{equation}
with the creation operators ${\alpha^{\pm\pm}}^\dagger(n_1,n_2)$
simply defined as the adjoint operators of $\alpha^{\pm\pm}(n_1,n_2)$.
One has
\begin{equation}
[\alpha^{\pm\pm}(n_1,n_2),{\alpha^{\pm\pm}}^\dagger(m_1,m_2)]=
\delta_{n_1,m_1}\delta_{n_2,m_2}\ \ \ , 
\end{equation}
while the annihilation (resp.  creation) operators clearly correspond, 
up to normalisation, to the Fourier modes of $\hat{A}_{\bar{z}}(z,\bar{z})$ 
(resp. $\hat{A}_z(z,\bar{z})$).

An overcomplete basis of the space of quantum states is then provided by
the coherent states,
\begin{equation}
|z^{\pm\pm}(n_1,n_2)>=\prod_{\pm\pm}\prod_{n_1,n_2}\
e^{-\frac{1}{2}|z^{\pm\pm}(n_1,n_2)|^2}\,
e^{z^{\pm\pm}(n_1,n_2)
{\alpha^{\pm\pm}}^\dagger(n_1,n_2)}\,|0>\ \ \ ,
\label{eq:coherent}
\end{equation}
with the following representation of the unit operator
\begin{equation}
\one=\int\,\prod_{\pm\pm}\prod_{n_1,n_2}
\frac{dz^{\pm\pm}(n_1,n_2)\wedge d\bar{z}^{\pm\pm}(n_1,n_2)}{\pi}\,
|z^{\pm\pm}(n_1,n_2)><z^{\pm\pm}(n_1,n_2)| \ \ \ ,
\end{equation}
where $z^{\pm\pm}(n_1,n_2)$ are arbitrary complex variables and $|0>$
is the usual Fock vacuum normalised such that $<0|0>=1$.
In particular, the gauge invariant physical states of the system are
those superpositions of these coherent states which are annihilated
by the first-class operator $\hat{\phi}(\vec{x})$, namely by all the
modes $\hat{\phi}^{\pm\pm}(n_1,n_2)$. However, this restriction does not
yet account for invariance under large gauge transformations,
which are not generated by the constraint operator $\hat{\phi}(\vec{x})$, 
a further specification to be addressed in the next section.

One could now proceed and solve for the physical state conditions
$\hat{\phi}(\vec{x})|\psi>=0$ in terms of the above mode decompositions.
However, we shall rather pursue the physical projector path, which will
enable us to solve these conditions by at the same time also determining the
wave functions of the corresponding physical states, and including the 
constraints which arise from the requirement of invariance under large
gauge transformations as well. Nonetheless, let us note that the
resolution of the physical state conditions $\hat{\phi}(\vec{x})|\psi>=0$
has been given in Ref.\cite{Bos1} precisely using the functional
coherent state representation of the algebra of the field degrees of freedom, 
to which we shall thus compare our results. The approach of Ref.\cite{Bos1,Bos2} 
however, uses the formal manipulation and resolution of the {\em functional\/}
differential equations expressing the physical state conditions in
the coherent state wave function representation of the
commutation relations for the field operators $\hat{A}_z(z,\bar{z})$ and 
$\hat{A}_{\bar{z}}(z,\bar{z})$. By working rather in terms of Fourier
modes as done here in the case of the torus $T_2$, such formal manipulations
are avoided by having only a {\em discrete\/} infinity of such operators, 
thus leaving only the much less critical issue of evaluating {\em discrete\/}
infinite products of normalisation factors for quantum states, 
for which $\zeta$-function regularisation techniques will be applied 
(since other regularisations would require some physical scale, hence
some geometry structure to be introduced on $T_2$).

\section{The Physical Projector}
\label{Sect5}

In order to construct the physical projector, which in effect projects out 
from any state its gauge variant components by averaging the state over all 
its gauge transformations and thereby leaving over its gauge invariant
components only\cite{Klauder1}, let us first consider the operator which induces
all finite small gauge transformations, namely 
$\hat{U}(\theta_0)=\exp\left(i/\hbar\int_{T_2}d^2\vec{x}\ \theta_0(\vec{x})
\hat{\phi}(\vec{x})\right)$. In terms of the previous mode representations
and definitions, one finds
\begin{equation}
\begin{array}{r l}
&-\frac{1}{4\pi}\sqrt{\frac{\tau_2}{\hbar N_k}}
\int_{T_2}d^2\vec{x}\,\theta_0(\vec{x})\hat{\phi}(\vec{x})=\\
 & \\
=&\frac{1}{2\sqrt{2}}\sum_{n_2=1}^{+\infty}
\Big[+n_2\theta_0^{+-}(0,n_2)\alpha^{++}(0,n_2)\Big]\ +\
\frac{1}{2\sqrt{2}}\sum_{n_1=1}^{+\infty}
\Big[-n_1\bar{\tau}\theta_0^{-+}(n_1,0)\alpha^{++}(n_1,0)\Big]\ +\ \\
 & \\
+&\frac{1}{2\sqrt{2}}\sum_{n_2=1}^{+\infty}
\Big[-n_2\theta_0^{++}(0,n_2)\alpha^{+-}(0,n_2)\Big]\ +\
\frac{1}{2\sqrt{2}}\sum_{n_1=1}^{+\infty}
\Big[+n_1\bar{\tau}\theta_0^{++}(n_1,0)\alpha^{-+}(n_1,0)\Big]\ +\ \\
 & \\
+&\frac{1}{4}\sum_{n_1=1}^{+\infty}\sum_{n_2=1}^{+\infty}
\Big[\Big(+n_2\theta_0^{+-}(n_1,n_2)-n_1\bar{\tau}\theta_0^{-+}(n_1,n_2)\Big)
\alpha^{++}(n_1,n_2)\Big]\ +\ \\
 & \\
+&\frac{1}{4}\sum_{n_1=1}^{+\infty}\sum_{n_2=1}^{+\infty}
\Big[\Big(-n_2\theta_0^{++}(n_1,n_2)-n_1\bar{\tau}\theta_0^{--}(n_1,n_2)\Big)
\alpha^{+-}(n_1,n_2)\Big]\ +\ \\
 & \\
+&\frac{1}{4}\sum_{n_1=1}^{+\infty}\sum_{n_2=1}^{+\infty}
\Big[\Big(+n_2\theta_0^{--}(n_1,n_2)+n_1\bar{\tau}\theta_0^{++}(n_1,n_2)\Big)
\alpha^{-+}(n_1,n_2)\Big]\ +\ \\
 & \\
+&\frac{1}{4}\sum_{n_1=1}^{+\infty}\sum_{n_2=1}^{+\infty}
\Big[\Big(-n_2\theta_0^{-+}(n_1,n_2)+n_1\bar{\tau}\theta_0^{+-}(n_1,n_2)\Big)
\alpha^{--}(n_1,n_2)\Big]\ +\ \Big[{\rm h.c.}\Big]\ \ \ .
\end{array}
\label{eq:gaugetransf}
\end{equation}

Note that, as it should, the zero mode $\theta_0^{++}(0,0)$ of the
gauge parameter function $\theta_0(\vec{x})$ does not appear in this
expression, and that the zero mode operators ${\alpha^{++}}^{(\dagger)}(0,0)$ do
not contribute either, showing once again that only small
gauge transformations are generated by the first-class constraint 
$\phi(\vec{x})$. Moreover, it is possible to verify that the commutators of 
the quantity in (\ref{eq:gaugetransf}) with the modes 
$\hat{A}_i^{\pm\pm}(n_1,n_2)$ do reproduce the expressions in 
(\ref{eq:gauge1}) while leaving the zero modes $\hat{A}_i^{++}(0,0)$ 
invariant, since we have, using the property in (\ref{eq:identity}),
\begin{equation}
\hat{U}(\theta_0)\,\hat{A}_i^{\pm\pm}(n_1,n_2)\,\hat{U}^{-1}(\theta_0)=
\hat{A}_i^{\pm\pm}(n_1,n_2)\,+\,
\Big[\frac{i}{\hbar}\int_{T_2}d^2\vec{x}\ \theta_0(\vec{x})
\hat{\phi}(\vec{x})\ ,\ \hat{A}_i^{\pm\pm}(n_1,n_2)\Big]\ \ \ .
\end{equation}

To determine how to construct the operator which generates the large
gauge transformations characterized by the holonomies $(k_1,k_2)$, let
us consider the zero mode $A_{\bar{z}}^{++}(0,0)$ (a similar analysis
is possible based on $A_z^{++}(0,0)$), whose gauge transformation is
given by (see (\ref{eq:gauge2}) and (\ref{eq:Az}))
\begin{equation}
\Delta A_{\bar{z}}^{++}(0,0)=-\frac{i}{2\tau_2}\ 
2\pi\Big[\tau k_1 - k_2\Big]\ \ \ .
\end{equation}
Consequently, the associated annihilation operator $\alpha^{++}(0,0)$
should transform according to
\begin{equation}
\hat{U}(k_1,k_2)\,\alpha^{++}(0,0)\,\hat{U}^{-1}(k_1,k_2)=\alpha^{++}(0,0)\ -\
2i\pi\sqrt{\frac{N_k}{\hbar\tau_2}}\,\Big[\tau k_1-k_2\Big]\ \ \ ,
\end{equation}
where $\hat{U}(k_1,k_2)$ stands for the operator generating the large
gauge transformation of holonomies $(k_1,k_2)$.
Therefore, we must have
\begin{equation}
\hat{U}(k_1,k_2)=C(k_1,k_2)\,
e^{\frac{2i\pi}{\hbar}\sqrt{\frac{\hbar N_k}{\tau_2}}
\Big\{\left[\bar{\tau}k_1-k_2\right]\alpha^{++}(0,0)+
\left[\tau k_1-k_2\right]{\alpha^{++}}^\dagger(0,0)\Big\}}\ \ \ ,
\end{equation}
where $C(k_1,k_2)$ is a cocycle factor to be determined presently such that
the group composition law is obeyed for large gauge transformations,
\begin{equation}
\hat{U}(\ell_1,\ell_2)\,\hat{U}(k_1,k_2)=\hat{U}(\ell_1+k_1,\ell_2+k_2)\ \ \ .
\end{equation}
Since---using the property in (\ref{eq:identity})---the latter constraint 
translates into the cocycle condition
\begin{equation}
e^{-4i\pi^2\frac{N_k}{\hbar}\left[\ell_1 k_2-\ell_2 k_1\right]}\,
C(\ell_1,\ell_2)\,C(k_1,k_2)=C(\ell_1+k_1,\ell_2+k_2)\ \ \ ,
\end{equation}
a careful analysis shows that the unique solution to this cocycle condition is
\begin{equation}
N_k=\frac{\hbar}{4\pi}\,k\ \ \ ,\ \ \ C(k_1,k_2)=e^{i\pi k k_1 k_2}\ \ \ ,
\label{eq:Nk}
\end{equation}
where $k=1,2,\dots$ is some positive integer value. This specific result
for the normalisation factor $N_k$ will thus be assumed henceforth.

These results therefore establish that
consistency of the quantised system under the action of large gauge
transformations in its zero mode sector $\hat{A}_i^{++}(0,0)$ requires the
quantisation of the normalisation factor $N_k$ in precisely such a manner
that a total of $k$ gauge invariant physical states are expected to
exist within the entire space of quantum states generated by the
coherent states constructed above. In addition, the operator
$\hat{U}(k_1,k_2)$ associated to the large gauge transformation of 
ho\-lo\-no\-my $(k_1,k_2)$ is thereby totaly specified, while the operator 
$\hat{U}(\theta_0)=\exp\left(i/\hbar\int_{T_2}d^2\vec{x}\,
\theta_0(\vec{x})\hat{\phi}(\vec{x})\right)$ associated to small gauge
transformations of parameter function $\theta_0(\vec{x})$ is defined
in (\ref{eq:gaugetransf}).

Consequently, the projector onto gauge invariant physical states should
simply be contructed by summing over all small and large gauge transformations
the action of the operators $\hat{U}(\theta_0)$ and $\hat{U}(k_1,k_2)$
just described. Even though this is straighforward for the large
gauge transformations, the summation over the small ones still requires
some further specifications\cite{Klauder1}, stemming from the fact that the 
spectrum of the non-zero modes of the gauge constraint $\hat{\phi}(\vec{x})=0$
is continuous. Since, as was seen previously, the small gauge invariance
is such that half the non-zero modes $A^{\pm\pm}_i(n_1,n_2)$ 
($n_1\ne$ or $n_2\ne 0$) may be set to zero, the other half being their
conjugate phase space variables, let us discuss this specific issue
in the following much simpler situation.

Consider a single degree of freedom system with coordinate $\hat{q}$ and
conjugate momentum $\hat{p}$, both hermitian operators, whose commutation
relation is the usual Heisenberg algebra $[\hat{q},\hat{p}]=i$,
and subject to the first-class constraint $\hat{q}=0$.
The spectrum of this latter operator being continuous, the proper definition
of a projector onto the states satisfying this constraint requires to
consider rather the projector onto those states whose $\hat{q}$ eigenvalues
lie within some interval $[-\delta,\delta]$, $\delta>0$ being a parameter
whose value may be as small as may be required\cite{Klauder1}. 
The latter projector is expressed as,
\begin{equation}
\proj\,_\delta\equiv\proj\,[-\delta<q<\delta]=\int_{-\delta}^{\delta}dq\,|q><q|=
\int_{-\infty}^{+\infty}d\xi\,e^{i\xi\hat{q}}\,\frac{\sin(\xi\delta)}{\pi\xi}
\ \ \ ,
\end{equation}
assuming that the position eigenstates are normalised such that
$<q_1|q_2>=\delta(q_1-q_2)$. By construction, one has the required properties,
\begin{equation}
\proj\,_\delta^2=\proj\,_\delta\ \ \ ,\ \ \ 
\proj\,_\delta^\dagger=\proj\,_\delta\ \ \ .
\end{equation}
However, one would rather wish to consider the operator singling out the
$|q=0>$ component of any state, namely,
\begin{equation}
\proj\,_0=|q=0><q=0|\ \ \ .
\end{equation}
Even though this operator is indeed hermitian, it is not strictly
in involution since one has,
\begin{equation}
\proj\,_0^2=\delta(0)\,\proj\,_0\ \ \ .
\end{equation}
In other words, since the position eigenstates of the $\hat{q}$ operators are 
non normalisable, the operator $\proj\,_0$ does not define a projection
operator in a strict sense, since even though it projects onto the $|q=0>$
component, it leads thereby to a non normalisable state satisfying the
constraint $\hat{q}=0$. Nevertheless, the non normalisable projector
$\proj\,_0$ may be constructed from the well defined one $\proj\,_\delta$
through the following limit,
\begin{equation}
\proj\,_0=\lim_{\delta\rightarrow 0}\,
\frac{1}{2\delta}\,\proj\,_\delta=\int_{-\infty}^{+\infty}\frac{d\xi}{2\pi}\,
e^{i\xi\hat{q}}\ \ \ .
\end{equation}
Hence, associated to the choice of normalisation of position eigenstates
$<q_1|q_2>=\delta(q_1-q_2)$, the non normalisable
projector $\proj\,_0$ onto the states such that $\hat{q}|\psi>=0$ is simply
represented by the integral operator on the r.h.s. of this last identity.

Transcribing these considerations to the U(1) Chern-Simons theory, it should 
be clear that the (non normalisable) projector onto the (non normalisable)
gauge invariant physical states of the system is simply given by (recall 
that the modes $\theta_0^{++}(0,0)$, $\theta_0^{+-}(n_1,0)$, 
$\theta_0^{-+}(0,n_2)$, $\theta_0^{--}(n_1,0)$ and $\theta_0^{--}(0,n_2)$ 
are non existent)
\begin{equation}
\proj\,_0=\sum_{k_1,k_2=-\infty}^{+\infty}\,\hat{U}(k_1,k_2)\,
\prod_{\pm\pm}\prod_{n_1,n_2}\int_{-\infty}^{+\infty}
\frac{d\theta_0^{\pm\pm}(n_1,n_2)}{2\pi}\,\hat{U}(\theta_0)\ \ \ .
\end{equation}
Note that this physical projector also defines the physical evolution
operator of the system, since the first-class Hamiltonian operator vanishes
identically, $\hat{H}=0$, a consequence of invariance under local coordinate
transformations in $\Sigma$.

\section{Gauge Invariant Physical States}
\label{Sect6}

The set of physical states of the system should now be identifiable
simply by applying the physical projector $\proj\,_0$ onto the entire
space of states representing the operator algebra of modes 
$\hat{A}^{\pm\pm}_i(n_1,n_2)$. Working in the basis (\ref{eq:coherent})
of coherent states $|z^{\pm\pm}(n_1,n_2)>$, this is tantamount
to considering the diagonal matrix elements
$<z^{\pm\pm}(n_1,n_2)|\proj\,_0|z^{\pm\pm}(n_1,n_2)>$, since these matrix
elements simply reduce to a sum of the physical state contributions as 
intermediate states, namely,
\begin{equation}
\begin{array}{r c l}
<z^{\pm\pm}(n_1,n_2)|\proj\,_0|z^{\pm\pm}(n_1,n_2)>&=&
\sum_{r}<z^{\pm\pm}(n_1,n_2)|r><r|z^{\pm\pm}(n_1,n_2)>\\
 & & \\
&=&\sum_r\,\left|<r|z^{\pm\pm}(n_1,n_2)>\right|^2\ \ \ ,
\end{array}
\label{eq:sumphysical}
\end{equation}
where $r$ is a discrete or continuous index labelling all
physical states (only a finite number $k$ of which are expected, of course).
Therefore, when having obtained an expression for the diagonal matrix elements
$<z^{\pm\pm}(n_1,n_2)|\proj\,_0|z^{\pm\pm}(n_1,n_2)>$ as a sum of modulus
squared quantities, the physical states of the system
together with their coherent state wave function representations are
readily identified up to a physically irrelevant phase factor, 
knowing that except for the factor $e^{-|z^{\pm\pm}(n_1,n_2)|^2/2}$ for 
each mode stemming from the normalisation of the coherent states, 
the wave functions $<r|z^{\pm\pm}(n_1,n_2)>$ are necessarily functions of 
the variables $z^{\pm\pm}(n_1,n_2)$ only, but not of their complex conjugate 
values $\bar{z}^{\pm\pm}(n_1,n_2)$. Had the system possessed some global 
symmetry beyond the U(1) gauge invariance, the associated different 
quantum numbers of the physical states could have used to label them, thereby 
making it easier to identify them as well as their wave functions from the above 
matrix elements suitably extended to include the action of the global symmetry
generators\cite{Klauder2}. Note also that trying to extract the same 
information---beginning with the number of physical states---from the 
partition function ${\rm Tr}\,\proj\,_0$ would be problematic. Indeed, that 
latter quantity is ill-defined since neither the physical states $|r>$ nor the 
physical projector $\proj\,_0$ are normalisable quantities.

Given that the exponential arguments appearing in the definition
of the gauge operators $\hat{U}(\theta_0)$ and $\hat{U}(k_1,k_2)$ are
linear in the creation and annihilation operators 
${\alpha^{\pm\pm}}^{(\dagger)}(n_1,n_2)$, the explicit evaluation of
the above diagonal coherent state matrix elements of the physical
projector $\proj\,_0$ is rather straighforward even though not immediate
for the modes with $n_1\ne 0$ and $n_2\ne 0$.
Let us only give the outline of the calculation for the modes
with $n_1=0$ or $n_2=0$. That for the modes with $n_1\ne 0$ and $n_2\ne 0$
is similar but involves some matrix algebra since the integrations
over $\theta_0^{++}(n_1,n_2)$ and $\theta_0^{--}(n_1,n_2)$ on the one
hand, and $\theta_0^{+-}(n_1,n_2)$ and $\theta_0^{-+}(n_1,n_2)$ on the
other, are coupled to one another in each case.

Consider again a single degree of freedom system with creation and
annihilation operators $a^\dagger$ and $a$, respectively,
such that $[a,a^\dagger]=1$, together with the associated coherent states
\begin{equation}
|z>=e^{-\frac{1}{2}|z|^2}\,e^{za^\dagger}\,|0>\ \ \ ,\ \ \ 
<z|z>=1\ \ \ ,
\end{equation}
$|0>$ being the usual Fock vacuum normalised as $<0|0>=1$. Using the identity
\begin{equation}
e^{A+B}=e^{-\frac{1}{2}[A,B]}\,e^A\,e^B\ \ \ ,
\label{eq:identity}
\end{equation}
valid for any two operators $A$ and $B$ which commute with their commutator
$[A,B]$, as well as the obvious properties that
\begin{equation}
a\,|z>=z\,|z>\ \ \ ,\ \ \ 
e^{i\lambda\theta a}\,|z>=e^{i\lambda\theta z}\,|z>\ \ \ ,
\end{equation}
where $\lambda$ and $\theta$ are arbitrary real and complex
variables, respectively (in keeping with
the notation in (\ref{eq:gaugetransf})),
one readily concludes that diagonal coherent state matrix elements
of an exponential operator whose argument is linear in the creation and
annihilation operators are simply given by
\begin{equation}
<z|e^{i\lambda[\theta a+\bar{\theta} a^\dagger]}|z>=
e^{-\frac{1}{2}\lambda^2|\theta|^2}\,e^{i\lambda\bar{\theta}\bar{z}}\,
e^{i\lambda\theta z}\ \ \ .
\label{eq:expoidentity}
\end{equation}

Consequently, in the case of the matrix elements
$<z^{\pm\pm}(n_1,n_2)|\proj\,_0|z^{\pm\pm}(n_1,n_2)>$,
the integrations over the discrete infinite number of 
non-zero modes $\theta_0^{\pm\pm}(n_1,n_2)$ ($n_1\ne 0$ or $n_2\ne 0$,
or both) for small gauge transformations simply correspond
to gaussian integrals for each of these contributions, whose results are then
multiplied with one another. As mentioned previously, the ensuing discrete
infinite products of normalisation factors are handled using 
$\zeta$-function re\-gu\-la\-risation techniques, rendering these products well 
defined. Moreover, the discrete infinite products of the remaining 
gaussian factors may be expressed as a simple exponential whose argument is 
given by the integral over the torus $T_2$ of a local quantity built from the
modes $z^{\pm\pm}(n_1,n_2)$ defining the coherent state 
$|z^{\pm\pm}(n_1,n_2)>$ for which the diagonal matrix element is evaluated.

Similar considerations apply to the contribution from the
zero mode sector $z^{++}(0,0)$, but the summation over the holonomies
$(k_1,k_2)$ is such that no clear separation of terms in the form of
(\ref{eq:sumphysical}), corresponding to the separate contributions of
the distinct physical states, appears to be feasable. In fact,
the evaluation of the zero mode sector contribution to the relevant
matrix elements requires an entirely different approach, which is
detailed in the Appendix and uses different representations of the
zero mode algebra (\ref{eq:commzero}). In particular, it is shown in the
Appendix that the total number of physical states is indeed equal
to the value of the integer $k$ which quantises the normalisation
factor $N_k$ in (\ref{eq:Nk}), so that the index $r$ introduced in 
(\ref{eq:sumphysical}) takes the following finite set of values,
$r=0,1,\dots,(k-1)$.

In order to give the final result of all these calculations in a convenient
form, let us introduce the following quantities associated to the
complex parameters $z^{\pm\pm}(n_1,n_2)$ (see (\ref{eq:Az})
and (\ref{eq:annihilation}))
\begin{equation}
A^{\pm\pm}_{\bar{z}}(n_1,n_2)=\sqrt{f^{\pm\pm}(n_1,n_2)\,\frac{4\pi}{k\tau_2}}\ \
z^{\pm\pm}(n_1,n_2)\ \ \ ,
\end{equation}
thereby determining a specific function $A_{\bar{z}}(z,\bar{z})$ through
its mode expansion.
In turn, let us then introduce the further modes defined in terms of those
of $A_{\bar{z}}(z,\bar{z})$,
\begin{equation}
\begin{array}{r c l}
\chi^{++}(n_1,n_2)&=&i\frac{\tau_2}{\pi}\,
\frac{-n_2A^{+-}_{\bar{z}}(n_1,n_2)-n_1\tau A^{-+}_{\bar{z}}(n_1,n_2)}
{n^2_1\tau^2-n^2_2}\ \ \ ,\\
 & & \\
\chi^{+-}(n_1,n_2)&=&i\frac{\tau_2}{\pi}\,
\frac{+n_2A^{++}_{\bar{z}}(n_1,n_2)-n_1\tau A^{--}_{\bar{z}}(n_1,n_2)}
{n^2_1\tau^2-n^2_2}\ \ \ ,\\
 & & \\
\chi^{-+}(n_1,n_2)&=&i\frac{\tau_2}{\pi}\,
\frac{+n_1\tau A^{++}_{\bar{z}}(n_1,n_2)-n_2 A^{--}_{\bar{z}}(n_1,n_2)}
{n^2_1\tau^2-n^2_2}\ \ \ ,\\
 & & \\
\chi^{--}(n_1,n_2)&=&i\frac{\tau_2}{\pi}\,
\frac{+n_1\tau A^{+-}_{\bar{z}}(n_1,n_2)+n_2 A^{-+}_{\bar{z}}(n_1,n_2)}
{n^2_1\tau^2-n^2_2}\ \ \ ,
\end{array}
\label{eq:chi}
\end{equation}
with again the understanding that the zero mode $\chi^{++}(0,0)=0$
is taken to vanish, and that the would-be non-zero modes $\chi^{+-}(n_1,0)$,
$\chi^{-+}(0,n_2)$, $\chi^{--}(n_1,0)$ and $\chi^{--}(0,n_2)$ 
($n_1,n_2=1,2,\dots$) do not
appear in the mode expansions. 

Correspondingly, the modes of the functions $\partial_{\bar{z}}\chi(z,\bar{z})$
and $\partial_z\chi(z,\bar{z})$ are such that on the one hand
\begin{equation}
\partial_{\bar{z}}\chi(z,\bar{z})=A_{\bar{z}}(z,\bar{z})-A_{\bar{z}}^{++}(0,0)
\ \ \ ,
\end{equation}
namely that all modes of $\partial_{\bar{z}}\chi(z,\bar{z})$ except for
the zero mode $\left(\partial_{\bar{z}}\chi\right)^{++}(0,0)=0$ coincide
with the corresponding ones of $A_{\bar{z}}(z,\bar{z})$, and on the other hand
the non-zero modes of $\partial_z\chi(z,\bar{z})$ are given by
\begin{equation}
\begin{array}{r c l}
\left(\partial_z\chi\right)^{++}(n_1,n_2)&=&
\frac{-1}{n^2_1\tau^2-n^2_2}\,\left\{\left[n^2_1|\tau|^2-n^2_2\right]
A_{\bar{z}}^{++}(n_1,n_2)+n_1n_2(\tau-\bar{\tau})A_{\bar{z}}^{--}(n_1,n_2)
\right\}\ \ \ ,\\
 & & \\
\left(\partial_z\chi\right)^{+-}(n_1,n_2)&=&
\frac{-1}{n^2_1\tau^2-n^2_2}\,\left\{\left[n^2_1|\tau|^2-n^2_2\right]
A_{\bar{z}}^{+-}(n_1,n_2)-n_1n_2(\tau-\bar{\tau})A_{\bar{z}}^{-+}(n_1,n_2)
\right\}\ \ \ ,\\
 & & \\
\left(\partial_z\chi\right)^{-+}(n_1,n_2)&=&
\frac{-1}{n^2_1\tau^2-n^2_2}\,\left\{-n_1n_2(\tau-\bar{\tau})
A_{\bar{z}}^{+-}(n_1,n_2)+\left[n^2_1|\tau|^2-n^2_2\right]
A_{\bar{z}}^{-+}(n_1,n_2)\right\}\ \ \ ,\\
 & & \\
\left(\partial_z\chi\right)^{--}(n_1,n_2)&=&
\frac{-1}{n^2_1\tau^2-n^2_2}\,\left\{+n_1n_2(\tau-\bar{\tau})
A_{\bar{z}}^{++}(n_1,n_2)+\left[n^2_1|\tau|^2-n^2_2\right]
A_{\bar{z}}^{--}(n_1,n_2)\right\}\ \ \ .
\end{array}
\end{equation}

An important identity that the modes of the function $\chi(z,\bar{z})$
satisfy is the functional relation
\begin{equation}
\partial_{\bar{z}}\left(\partial_z\chi\right)=
\partial_z A_{\bar{z}}=\partial_z\left(\partial_{\bar{z}}\chi\right)\ \ \ .
\label{eq:identitychi}
\end{equation}

In terms of these different quantities, finally the coherent state
wave function for each of the $k$ physical states $|r>$ of the system is given
by ($r=0,1,2,\dots,k-1$),
\begin{equation}
\begin{array}{r c l}
<r|A_{\bar{z}}(z,\bar{z})>&\equiv&<r|z^{\pm\pm}(n_1,n_2)>\\
& & \\
&=&e^{-\frac{k\tau_2}{2\pi}\left(A^{++}_{\bar{z}}(0,0)\right)^2}\ 
\frac{1}{\eta(\tau)}\ \Theta\left[\begin{array}{c}
	r/k\\
	0
	\end{array}\right]
\Big(-i\frac{k\tau_2}{\pi}A^{++}_{\bar{z}}(0,0)\Big|k\tau\Big)\times\\
 & & \\
&\times&\,e^{-\frac{ik}{4\pi}\int_{T_2}dz\wedge d\bar{z}\,
\left|A_{\bar{z}}(z,\bar{z})\right|^2}\,
e^{\frac{ik}{4\pi}\int_{T_2}dz\wedge d\bar{z}\,
\partial_{\bar{z}}\chi(z,\bar{z})\partial_z\chi(z,\bar{z})}\ \ \ ,
\label{eq:wavefunctions}
\end{array}
\end{equation}
where $\eta(\tau)$ is the Dedekind $\eta$-function,
$\eta=e^{i\pi\tau/12}\prod_{n=1}^{+\infty}\left(1-e^{2i\pi n\tau}\right)$,
$\Theta$ is the torus $\theta$-function with characteristics whose
definition is given in the Appendix, while the
physically irrelevant overall phase factor is set to unity.
Note that in the last two exponential factors, the integral
$\int_{T_2}dz\wedge d\bar{z}\left|A_{\bar{z}}(z,\bar{z})\right|^2$
does include the contribution\ \ 
${\rm exp}\left(-k\tau_2|A_{\bar{z}}^{++}(0,0)|^2/(2\pi)\right)$
from the zero mode component $A_{\bar{z}}^{++}(0,0)$
of the coherent state $|z^{\pm\pm}(n_1,n_2)>$, in contradistinction to the
second integral $\int_{T_2}dz\wedge d\bar{z}
\partial_{\bar{z}}\chi(z,\bar{z})\partial_z\chi(z,\bar{z})$ which only
includes contributions from the non-zero modes $A_{\bar{z}}^{\pm\pm}(n_1,n_2)$
($n_1\ne 0$ or $n_2\ne 0$). This point is important when checking gauge and 
modular invariance properties of these physical wave functions.

These wave functions coincide with those established in Ref.\cite{Bos1}
by using a functional representation of the commutations relations
of the field degrees of freedom $A_i(\vec{x})$ in order to solve for the
physical state condition $\hat{\phi}(\vec{x})=0$ as well as requiring
invariance under large gauge transformations. This identity of
results thus demonstrates that the physical projector approach is indeed
capable, without any gauge fixing procedure whatsoever and thereby avoiding
the possibility of any Gribov problem, to properly identify the actual gauge
invariant content of a system, even when this requires the projection down
to only a finite number of components from an initially infinite number
of states.

By construction, the states whose coherent state wave functions are
given in (\ref{eq:wavefunctions}) are invariant under both large and 
small gauge transformations. However, this property does not necessarily
imply that the wave functions themselves are invariant under arbitrary
gauge transformations of the variables $A_{\bar{z}}(z,\bar{z})$.
To make this point explicit, let us consider the gauge transformations
of the physical states, whose invariance should thus imply the following
identities, 
\begin{equation}
\begin{array}{r c l}
<r|\hat{U}^{-1}(\theta_0)|A_{\bar{z}}(z,\bar{z})>&=&
<r|A_{\bar{z}}(z,\bar{z})>\ \ \ ,\\
 & & \\
<r|\hat{U}^{-1}(k_1,k_2)|A_{\bar{z}}(z,\bar{z})>&=&
<r|A_{\bar{z}}(z,\bar{z})>\ \ \ ,
\end{array}
\label{eq:gaugeinvariance}
\end{equation}
valid both for small and large gauge transformations $\hat{U}(\theta_0)$
and $\hat{U}(k_1,k_2)$
whatever the values of their modes $\theta_0^{\pm\pm}(n_1,n_2)$ or their
holonomies $(k_1,k_2)$. 

In the case of small gauge transformations, a careful analysis of the
contributions to the relevant matrix elements establishes the result,
\begin{equation}
\hat{U}^{-1}(\theta_0)|A_{\bar{z}}(z,\bar{z})>=
e^{-\frac{ik}{4\pi}\int_{T_2}dz\wedge d\bar{z}\left[
\partial_{\bar{z}}\chi\partial_z\theta_0-
\partial_{\bar{z}}\theta_0\overline{\partial_{\bar{z}}\chi}\,\right]}\
|A_{\bar{z}}(z,\bar{z})+\partial_{\bar{z}}\theta_0(z,\bar{z})>\ \ \ ,
\end{equation}
as well as
\begin{equation}
<r|A_{\bar{z}}(z,\bar{z})+\partial_{\bar{z}}\theta_0(z,\bar{z})>=
e^{-\frac{ik}{4\pi}\int_{T_2}dz\wedge d\bar{z}\left[
(A_{\bar{z}}-\partial_{\bar{z}}\chi)\partial_z\theta_0+
\partial_{\bar{z}}\theta_0(\overline{A_{\bar{z}}}-\partial_z\chi)\right]}\
<r|A_{\bar{z}}(z,\bar{z})>\ \ \ ,
\end{equation}
where $A_{\bar{z}}(z,\bar{z})+\partial_{\bar{z}}\chi(z,\bar{z})$ stands for
the transformations of the modes $A_{\bar{z}}^{\pm\pm}(n_1,n_2)$ under
the small gauge transformation $\theta_0(\vec{x})$ of modes
$\theta_0^{\pm\pm}(n_1,n_2)$, with of course the understanding
that the zero mode $A_{\bar{z}}^{++}(0,0)$ is left invariant.

Combining these two relations and using the identity in 
(\ref{eq:identitychi}) after integration by parts in the
exponential factors (to which the zero mode $A^{++}_{\bar{z}}(0,0)$
does not contribute), the first identity in (\ref{eq:gaugeinvariance})
then indeed follows, thereby confirming gauge invariance of the physical
states under small gauge transformations.

Similarly for a large gauge transformation $\hat{U}(k_1,k_2)$ of holonomies
$(k_1,k_2)$, one finds,
\begin{equation}
\begin{array}{r l}
\hat{U}^{-1}(k_1,k_2)&|A_{\bar{z}}(z,\bar{z})>=\\
& \\
=&e^{i\pi kk_1k_2}\,
e^{-\frac{1}{2}ik\left[(k_1\tau-k_2)\overline{A_{\bar{z}}^{++}(0,0)}+
(k_1\bar{\tau}-k_2)A_{\bar{z}}^{++}(0,0)\right]}\
|A_{\bar{z}}(z,\bar{z})-\frac{i\pi}{\tau_2}(\tau k_1-k_2)>\ \ \ ,
\end{array}
\end{equation}
as well as
\begin{equation}
<r|A_{\bar{z}}(z,\bar{z})-\frac{i\pi}{\tau_2}(\tau k_1-k_2)>=
e^{-i\pi kk_1k_2}\,
e^{\frac{1}{2}ik\left[(k_1\tau-k_2)\overline{A_{\bar{z}}^{++}(0,0)}+
(k_1\bar{\tau}-k_2)A_{\bar{z}}^{++}(0,0)\right]}\
<r|A_{\bar{z}}(z,\bar{z})>\ \ \ ,
\end{equation}
where this time $A_{\bar{z}}(z,\bar{z})-i\pi(\tau k_1-k_2)/\tau_2$
stands for the transformation of the modes $A_{\bar{z}}^{\pm\pm}(n_1,n_2)$
under the large gauge transformation of holonomies $(k_1,k_2)$ which
of course affects only the zero mode $A_{\bar{z}}^{++}(0,0)$ by the
indicated constant shift linear in $k_1$ and $k_2$.
Consequently, these two relations also lead to the
the second identity in (\ref{eq:gaugeinvariance}), thereby establishing
gauge invariance of the physical states under large gauge transformations
as well.

These relations also demonstrate that in spite of
the gauge invariance of the physical states $|r>$ $(r=0,1,\dots,k-1)$ under
small and large transformations,
$\hat{U}(\theta_0)|r>=|r>$ and $\hat{U}(k_1,k_2)|r>=|r>$,
their coherent state wave functions $<r|A_{\bar{z}}(z,\bar{z})>$ are not 
invariant by themselves under the simple substitution of the
gauge variation of their argument $A_{\bar{z}}(z,\bar{z})$, namely 
the variations $A_{\bar{z}}'(z,\bar{z})=A_{\bar{z}}(z,\bar{z})+
\partial_{\bar{z}}\chi(z,\bar{z})$ for small gauge transformations and
$A_{\bar{z}}'(z,\bar{z})=A_{\bar{z}}(z,\bar{z})-i\pi(\tau k_1-k_2)/\tau_2$
for large ones.

Let us now turn to the issue of modular invariance of the physical content
of the quantised system. If indeed this content depends only on the
complex structure---parametrised by the variable $\tau$---as the only
structure necessary beyond the mere topological one of the underlying
torus $T_2$, the spectrum of physical states should remain invariant
under the modular group $PSL(2,Z\!\!\!Z)$ of $T_2$
which is generated by the transformations
\begin{equation}
T\ :\ \ \ \tau\rightarrow\tau+1\ \ \ ,\ \ \ 
S\ :\ \ \ \tau\rightarrow -\frac{1}{\tau}\ \ \ .
\end{equation}
Indeed, these transformations correspond to global diffeomorphisms---namely
Dehn twists---in the local trivialisation of $T_2$ characterized through 
the choice of holonomy basis ($a$,$b$) and the associated
coordinates $0<x^1,x^2<1$, and these modular transformations
define equivalences classes for the values of $\tau$ which correspond 
to a same complex structure (or conformal class) on $T_2$.

An explicit analysis of the transformation properties of the physical
wave functions (\ref{eq:wavefunctions}) shows that the requirement of
invariance of the physical content of the system under the $T$ mo\-du\-lar
transformation is met only if the integer $k$ {\em also takes an even value\/}
(see the Appendix)\cite{Bos1}, in which case one finds,
\begin{equation}
T\ :\ \ \ <r|A_{\bar{z}}(z,\bar{z})>\ \rightarrow\  e^{-i\pi/12}\,
e^{i\pi r^2/k}\,<r|A_{\bar{z}}(z,\bar{z})>\ \ \ .
\end{equation}
On the other hand, invariance under the $S$ modular transformation
is realised through the transformations,
\begin{equation}
S\ :\ \ \ \tau\ \rightarrow\ \tilde{\tau}=-\frac{1}{\tau}\ \ \ ,\ \ \ 
A_{\bar{z}}^{++}(0,0)\ \rightarrow\  \tilde{A}_{\bar{z}}^{++}(0,0)=
-\bar{\tau}A_{\bar{z}}^{++}(0,0)\ \ \ ,
\end{equation}
while the non-zero modes $A_{\bar{z}}^{\pm\pm}(n_1,n_2)$ are left
unchanged, in which case one has (see the Appendix),
\begin{equation}
S\ :\ \ \ <r|A_{\bar{z}}(z,\bar{z})>_{\tau}\ \rightarrow\
<r|\tilde{A}_{\bar{z}}(z,\bar{z})>_{\tilde{\tau}}\ =
\sum_{r'=0}^{k-1}\,\frac{1}{\sqrt{k}}\,
e^{2i\pi rr'/k}\,<r'|A_{\bar{z}}(z,\bar{z})>_{\tau}\ \ \ .
\end{equation}

Hence, in addition to the quantisation condition $N_k=\hbar k/(4\pi)$
imposed on the normalisation factor $N_k$ by the requirement of
invariance under large gauge transformations, modular invariance of 
the theory requires also that the integer $k$ be even\cite{Bos1}. In this case 
the space of quantum physical states 
does provide an irreducible representation of the $T_2$ modular group, 
showing that the physical content of the system does indeed depend only on 
the choice of complex structure (or conformal class) on $T_2$ characterized 
through the equivalence class under the modular group of the parameter $\tau$.
Nevertheless, physical states are not individually modular invariant, 
but rather they define a projective representation of the
modular group. This is how close the quantised U(1) Chern-Simons theory
is to being a purely topological quantum field theory, the dependency
on a complex structure on $T_2$ following from the existence of a
conformal anomaly at the quantum level\cite{Witten2}.

\section{Conclusions}
\label{Sect7}

This paper has demonstrated that the new approach to the quantisation
of gauge invariant systems\cite{Klauder1}, based on the physical projector 
onto the subspace of gauge invariant states, is perfectly adequate to handle the
intricacies of topological quantum fields theories, in which only a finite
set of physical states is to remain after the infinity of gauge variant
configurations has been projected away. This new approach to
the quantisation of constrained systems does not require any gauge 
fixing procedure whatsoever and is thus free of any potential Gribov 
ambiguity in the case of gauge symmetries\cite{Gov2}, 
in contradistinction to all other
quantisation frameworks for gauge invariant systems. Another of its
advantages is that the physical projector approach to the quantisation
of such theories is directly set simply within Dirac's formulation, 
where it finds its natural place.
In particular, the physical projector enables the construction of the
physical evolution operator of such systems, to which only physical
states contribute as intermediate states, so that the physical content may
be identified directly from the matrix elements of that evolution
operator, including the wave functions of physical states. 
For topological field theories, since these systems are 
invariant under local diffeomorphisms in the base manifold, 
their gauge invariant Hamiltonian vanishes identically, in
which case the evolution operator coincides with the physical projector.

More specifically, the physical projector approach was applied
to the U(1) pure Chern-Simons theory in 2+1 dimensions in a space
whose topology is that of $I\!\!R\times T_2$, $T_2$ being an arbitrary
two dimensional torus. Through a careful analysis of the quantised system,
of the relevant physical projector, and in particular of the specific
discrete infinite mode content of the system, it has been possible to identify 
and construct the physical spectrum and the co\-he\-rent state wave functions of 
the gauge invariant states, leading to results which are
in complete agreement with those of other approaches to the quantisation 
of the same system\cite{Witten2,Bos1,Bos2,Lab,CS}, while also avoiding 
some of complications or formal manipulations inherent to these other 
approaches. The only more or less ad-hoc but unavoidable feature which 
is introduced in our ana\-ly\-sis is that the discrete infinite products---rather 
than continuous ones as occurs in functional representations---of gaussian 
normalisation factors have been evaluated using $\zeta$-function 
regu\-la\-ri\-sa\-tion, which avoids having to introduce any other structure on 
the underlying two dimensional Riemann surface beyond those associated already
to the topology and complex structure (or conformal class) of that surface.
This is in keeping with the fact that the quantised theory only depends
on that complex structure but no other structure beyond it (even when one
is introduced, namely through a metric structurec\cite{Witten2}),
while also a dependency on the complex structure rather than purely
the topology of the underlying manifold is the unavoidable consequence
of the quantisation of the system\cite{Witten2}. In particular, it was shown how
gauge invariance under large gauge transformations implies a quantisation
rule for the overall normalisation of the classical Chern-Simons action
in terms of an integer equal to the number of physical states,
which in turn is also required to be even for modular
invariance to be realised\cite{Bos1}. When both these restrictions are met,
the quantised system indeed only depends on the complex structure introduced 
on the underlying Riemann surface.

These systems are also distinguished by the fact that their physical phase
space is a compact manifold, in the present instance with the topology
of a two dimensional torus, which is in contradistinction to the ordinary
situation in which the phase space of a given system is a cotangent
bundle. Usually, geometric quantisation techniques are then invoked
in order to address the specific issues raised by a phase space having
a compact topology\cite{Witten3}. Nevertheless, no such techniques were 
introduced here, but rather by properly identifying the operator which generates
the transformations responsible for such a compact topology of phase
space---in the present instance large gauge transformations---, it was
possible to properly represent the consequences of such a circumstance
using straightforward coherent state techniques of ordinary quantum mechanics.
Clearly, similar considerations
based on the construction of the relevant projection operator are of
application to any system whose phase space includes a compact ma\-ni\-fold
which is an homogeneous coset space $G/H$, $G$ and $H$ being compact
Lie algebras. From that point of view, it may well be worthwhile
to explore the potential of the physical projector as an
alternative to geometric quantisation techniques.

The physical projector approach is thus quite an efficient
approach to the quantisation of gauge invariant systems, which does not
require any gauge fixing procedure whatsoever and thus avoids the potential
Gribov problems inherent to such procedures. Some of its advantages have
been illustrated here in the instance of the U(1) Chern-Simons
theory, as well as for some simple gauge invariant quantum mechanical
systems elsewhere\cite{Klauder2,Villa}. Hence, it appears timely now 
to start exploring the application\cite{Klauder3} of this alternative method 
to the quantisation of gauge invariant systems of more direct physical 
interest, within the context of the recent developments surrounding M-theory
compactified to low dimensions, and aiming beyond that
towards the gauge invariant theories of the fundamental interactions among 
the elementary quantum excitations in the natural Universe.

\section*{Acknowledgments}

Prof. John Klauder is gratefully acknowledged for useful conversations
concerning the physical projector, and for his
constant interest in this work. This work is part of the undergraduate Diploma 
Thesis of B.D.

\clearpage

\section*{Appendix}

This Appendix outlines the calculation of the contribution of the
zero mode sector to the coherent state diagonal matrix elements
$<z^{\pm\pm}(n_1,n_2)|\proj\,_0|z^{\pm\pm}(n_1,n_2)>$ of the
physical projector operator, namely the quantity
\begin{equation}
\sum_{k_1,k_2=-\infty}^{+\infty}\,<z|\hat{U}(k_1,k_2)|z>=
\sum_{k_1,k_2=-\infty}^{+\infty}\ <z|e^{i\pi kk_1k_2}\,
e^{i\sqrt{\frac{\pi k}{\tau_2}}\left\{[\bar{\tau}k_1-k_2]\alpha+
[\tau k_1-k_2]\alpha^\dagger\right\}}|z>\ \ \ .
\label{eq:MEzero}
\end{equation}
Here, $z$, $\alpha$ and $\alpha^\dagger$ stand of course for the
zero mode sector only, with the obvious understanding that usual
indices distinguishing these zero modes, such as $z^{++}(0,0)$, 
are not explicitely displayed in this Appendix.

Since the evaluation of these matrix elements requires changes of bases
for different re\-pre\-sen\-ta\-tions of the associated quantum algebra,
let us recall here the relations between the different quantum operators
appearing in this sector of the system. One has the definitions
\begin{equation}
\alpha=\frac{1}{2}\sqrt{\frac{k}{\pi\tau_2}}\,
\left[-i\tau \hat{A}_1+i\hat{A}_2\right]\ \ \ ,\ \ \
\alpha^\dagger=\frac{1}{2}\sqrt{\frac{k}{\pi\tau_2}}\,
\left[i\bar{\tau} \hat{A}_1-i\hat{A}_2\right]\ \ \ ,\ \ \
\end{equation}
while the corresponding commutation rules are
\begin{equation}
\left[\hat{A}_1,\hat{A}_2\right]=\frac{2i\pi}{k}\ \ \ ,\ \ \ 
\left[\alpha,\alpha^\dagger\right]=1\ \ \ ,
\label{eq:commutation}
\end{equation}
with the implicit understanding that both $\hat{A}_1$ and $\hat{A}_2$ are
hermitian operators. Using these relations as well as the identity in 
(\ref{eq:identity}), large gauge transformations also read as
\begin{equation}
\hat{U}(k_1,k_2)=\ e^{i\pi kk_1k_2}\,
e^{ik\left[k_1\hat{A}_2-k_2\hat{A}_1\right]}=
\ e^{ikk_1\hat{A}_2}\,e^{-ikk_2\hat{A}_1}
\ \ \ .
\label{eq:EA1A2}
\end{equation}

Since the algebra for the modes $\hat{A}_1$ and $\hat{A}_2$ is that
of the usual Heisenberg algebra, with $\hat{A}_1$ playing the r\^ole of
the configuration space coordinate and $\hat{A}_2$ that of the conjugate
momentum variable, it is clear from the above last expression of the
operator $\hat{U}(k_1,k_2)$ in terms of these modes that the mixed
configuration-momentum space matrix elements of $\hat{U}(k_1,k_2)$ 
are readily obtained. Hence, let us develop the different representations
of the commutations relations (\ref{eq:commutation}), namely the
configuration space, the momentum space and the coherent state ones.

Configuration and momentum space representations correspond to
eigenstates $|A_1>$ and $|A_2>$ of the $\hat{A}_1$ and $\hat{A}_2$ operators, 
respectively,
\begin{equation}
\hat{A}_1|A_1>=A_1|A_1>\ \ \ ,\ \ \ 
\hat{A}_2|A_2>=A_2|A_2>\ \ \ ,
\end{equation}
whose normalisation is chosen to be such that
\begin{equation}
<A_1|A'_1>=\delta(A_1-A'_1)\ \ \ ,\ \ \ 
<A_2|A'_2>=\delta(A_2-A'_2)\ \ \ ,
\end{equation}
to which the following representations of the identity operator are thus
associated
\begin{equation}
\one=\int_{-\infty}^{+\infty}dA_1\,|A_1><A_1|\ \ \ ,\ \ \ 
\one=\int_{-\infty}^{+\infty}dA_2\,|A_2><A_2|\ \ \ .
\end{equation}

Since the factor $\hbar'=2\pi/k$ plays the r\^ole of an effective Planck 
constant, it is clear that the configuration space wave function
representations of the two operators $\hat{A}_1$ and $\hat{A}_2$ are simply,
\begin{equation}
<A_1|\hat{A}_1|\psi>=A_1\,<A_1|\psi>\ \ \ ,\ \ \ 
<A_1|\hat{A}_2|\psi>=-\frac{2i\pi}{k}\,\frac{\partial}{\partial A_1}\,
<A_1|\psi>\ \ \ ,
\label{eq:repA1A2}
\end{equation}
and for the momentum space wave function representations,
\begin{equation}
<A_2|\hat{A}_1|\psi>=\frac{2i\pi}{k}\,\frac{\partial}{\partial A_2}\,
<A_2|\psi>\ \ \ ,\ \ \ 
<A_2|\hat{A}_2|\psi>=A_2\,<A_2|\psi>\ \ \ .
\end{equation}
In particular, given the above choice of normalisation, we have for the
matrix elements expressing the corresponding changes of basis,
\begin{equation}
<A_2|A_1>=\frac{\sqrt{k}}{2\pi}\,e^{-\frac{ik}{2\pi}A_1A_2}\ \ \ ,\ \ \ 
<A_1|A_2>=\frac{\sqrt{k}}{2\pi}\,e^{+\frac{ik}{2\pi}A_1A_2}\ \ \ .
\end{equation}

Let us now consider the Fock state representation of the same quantum
algebra, whose set of orthonormalised basis vectors is thus defined by
\begin{equation}
|n>=\frac{1}{\sqrt{n!}}\,\left(\alpha^\dagger\right)^n\,|0>\ \ \ ,
\end{equation}
$|0>$ being of course the Fock vacuum normalised such that
$<0|0>=1$. Given the above configuration space representation,
the configuration space wave functions of the Fock basis vectors
are easily constructed. For the vacuum, one finds from the condition
$\alpha|0>=0$, including proper normalisation,
\begin{equation}
<A_1|0>=\left(\frac{k\tau_2}{2\pi^2}\right)^{1/4}\,
e^{\frac{ik}{4\pi}\tau A_1^2}\ \ \ ,
\end{equation}
while the excited Fock states are such that
\begin{equation}
<A_1|n>=\left(\frac{k\tau_2}{2\pi^2}\right)^{1/4}\,\frac{1}{\sqrt{n!}}\,
\left(\frac{i\bar{\tau}}{2\tau_2}\right)^{n/2}\,
\left[u-\frac{d}{du}\right]^n\,e^{\frac{1}{2}\frac{\tau}{\bar{\tau}}u^2}
\ \ \ ,\ \ \ u=\sqrt{\frac{ik\bar{\tau}}{2\pi}}\,A_1\ \ \ .
\end{equation}
In order to solve for these latter expressions, let us introduce
polynomials $P_n(u;\lambda)$ generalising the usual Hermite polynomials,
and defined by the generating function
\begin{equation}
e^{-\frac{1}{2}(1+\lambda)t^2+(1+\lambda)tu}=\sum_{n=0}^{+\infty}\,
\frac{t^n}{n!}\,P_n(u;\lambda)\ \ \ ,
\end{equation}
where $\lambda$ is a complex parameter.
These polynomials satisfy the following set of properties,
\begin{equation}
P_n(u;\lambda)=e^{\frac{1}{2}\lambda u^2}\,
\left[u-\frac{d}{du}\right]^n\,e^{-\frac{1}{2}\lambda u^2}\ \ \ ,
\end{equation}
\begin{equation}
P_{n+1}(u;\lambda)=(1+\lambda)uP_n(u;\lambda)-\frac{d}{du}P_n(u;\lambda)\ \ \ ,
\end{equation}
\begin{equation}
P_0(u;\lambda)=1\ \ \ ,\ \ \ 
P_1(u;\lambda)=(1+\lambda)u\ \ \ .
\end{equation}
Hence finally, one finds
\begin{equation}
<A_1|n>=\left(\frac{k\tau_2}{2\pi^2}\right)^{1/4}\,\frac{1}{\sqrt{n!}}\,
\left(\frac{i\bar{\tau}}{2\tau_2}\right)^{n/2}\,
e^{\frac{ik\tau}{4\pi}A_1^2}\,
P_n\left(\sqrt{\frac{ik\bar{\tau}}{2\pi}}\,A_1;
\frac{-i\tau}{i\bar{\tau}}\right)\ \ \ .
\end{equation}

Similarly for the momentum wave functions, one has
\begin{equation}
<A_2|n>=\left(\frac{k\tau_2}{2\pi^2|\tau|^2}\right)^{1/4}\,
\frac{(-i)^n}{\sqrt{n!}}\,
\left(\frac{i\bar{\tau}}{2\tau_2}\right)^{n/2}\,
e^{\frac{k}{4i\pi\tau}A_2^2}\,
P_n\left(\sqrt{\frac{k}{2i\pi\bar{\tau}}}\,A_2;
\frac{i\bar{\tau}}{-i\tau}\right)\ \ \ .
\end{equation}

Finally, let us consider the coherent state basis,
\begin{equation}
|z>=e^{-\frac{1}{2}|z|^2}\,e^{z\alpha^\dagger}\,|0>\ \ \ ,\ \ \ 
\one=\int\frac{dz\wedge d\bar{z}}{\pi}\,|z><z|\ \ \ .
\end{equation}
Using the relations
\begin{equation}
\alpha|z>=z|z>\ \ \ ,\ \ \ 
<n|z>=\frac{z^n}{\sqrt{n!}}\,e^{-\frac{1}{2}|z|^2}\ \ \ ,
\end{equation}
as well as the above generating function for the polynomials $P_n(u;\lambda)$
which appear in the matrix elements $<A_1|n>$ and $<A_2|n>$, the following
results are readily obtained,
\begin{equation}
<A_1|z>=\left(\frac{k\tau_2}{2\pi^2}\right)^{1/4}\,
e^{-\frac{1}{2}|z|^2}\,e^{\frac{ik\tau}{4\pi}A_1^2}\,
e^{-\frac{1}{2}z^2+zA_1\sqrt{\frac{k\tau_2}{\pi}}}\ \ \ ,
\end{equation}
\begin{equation}
<A_2|z>=\left(\frac{k\tau_2}{2\pi^2|\tau|^2}\right)^{1/4}\,
e^{-\frac{1}{2}|z|^2}\,e^{\frac{k}{4i\pi\tau}A_2^2}\,
e^{\frac{1}{2}\frac{i\bar{\tau}}{-i\tau}z^2+\frac{zA_2}{\tau}
\sqrt{\frac{k\tau_2}{\pi}}}\ \ \ .
\end{equation}

Having established these different changes of bases, let us return to the
evaluation of the matrix element (\ref{eq:MEzero}). Obviously,
given (\ref{eq:EA1A2}), the mixed configuration-momentum space
matrix elements are simply,
\begin{equation}
<A_2|\sum_{k_1,k_2=-\infty}^{+\infty}\hat{U}(k_1,k_2)|A_1>=
\sum_{k_1,k_2=-\infty}^{+\infty}\,
\frac{\sqrt{k}}{2\pi}\,e^{ikk_1A_2}\,e^{-ikk_2A_1}\,e^{-\frac{ik}{2\pi}A_1A_2}
\ \ \ .
\end{equation}
Using the identities
\begin{equation}
\sum_{k_1=-\infty}^{+\infty}e^{ikk_1A_2}=\sum_{n_1=-\infty}^{+\infty}\,
\frac{2\pi}{k}\delta\left(A_2-\frac{2\pi n_1}{k}\right)\ \ \ ,
\end{equation}
\begin{equation}
\sum_{k_2=-\infty}^{+\infty}e^{-ikk_2A_1}=\sum_{n_2=-\infty}^{+\infty}\,
\frac{2\pi}{k}\delta\left(A_1-\frac{2\pi n_2}{k}\right)\ \ \ ,
\end{equation}
one also has
\begin{equation}
<A_2|\sum_{k_1,k_2=-\infty}^{+\infty}\hat{U}(k_1,k_2)|A_1>=
\frac{2\pi}{k}\,\frac{1}{\sqrt{k}}\,\sum_{n_1,n_2=-\infty}^{+\infty}\,
e^{-\frac{2i\pi}{k}n_1n_2}\,\delta\left(A_2-\frac{2\pi n_1}{k}\right)
\,\delta\left(A_1-\frac{2\pi n_2}{k}\right)\ \ \ .
\end{equation}

However, in this form it is not yet possible to identify the different
contributions of the physical states in the form of modulus squared
terms. In order to achieve that aim,
let us finally compute the configuration space matrix elements of the
relevant operator, using the change of basis $<A_1|A_2>$ above. One
then obtains,
\begin{equation}
\begin{array}{r l}
&<A_1|\sum_{k_1,k_2=-\infty}^{+\infty}\hat{U}(k_1,k_2)|A_1'>=
\int_{-\infty}^{+\infty}dA_2\,<A_1|A_2>
<A_2|\sum_{k_1,k_2=-\infty}^{+\infty}\hat{U}(k_1,k_2)|A_1'>\\
 & \\
=&\frac{2\pi}{k}\sum_{n_2=-\infty}^{+\infty}
\left[\delta\left(A_1'-\frac{2\pi n_2}{k}\right)\
\sum_{n_1=-\infty}^{+\infty}\,\delta\left(A_1-\frac{2\pi n_2}{k}-2\pi n_1\right)
\right]\ \ \ .
\end{array}
\end{equation}

Having thus obtained the configuration space matrix elements of the
zero mode physical projector operator, let us finally apply it onto
any state of this sector of the quantised system. Introducing the
configuration space wave function
\begin{equation}
|\psi>=\int_{-\infty}^{+\infty}dA_1\,|A_1><A_1|\psi>=
\int_{-\infty}^{+\infty}dA_1\,|A_1>\,\psi(A_1)\ \ \ ,\ \ \ 
\psi(A_1)\equiv <A_1|\psi>\ \ \ ,
\end{equation}
one readily derives,
\begin{equation}
\begin{array}{r l}
&<A_1|\sum_{k_1,k_2=-\infty}^{+\infty}\hat{U}(k_1,k_2)|\psi>=
\int_{-\infty}^{+\infty}dA_1'\,
<A_1|\sum_{k_1,k_2=-\infty}^{+\infty}\hat{U}(k_1,k_2)|A_1'><A_1'|\psi>\\
 & \\
=&\frac{2\pi}{k}\sum_{n_2=-\infty}^{+\infty}\,
\left[\psi\left(\frac{2\pi n_2}{k}\right)\,
\sum_{n_1=-\infty}^{+\infty}\delta\left(A_1-\frac{2\pi n_2}{k}-2\pi n_1\right)
\right]\ \ \ .
\end{array}
\end{equation}
In particular, since physical states are to be invariant under the action of
the physical projector $\sum_{k_1,k_2=-\infty}^{+\infty}\hat{U}(k_1,k_2)$,
they should thus obey the following equation,
\begin{equation}
\frac{2\pi}{k}\sum_{n_2=-\infty}^{+\infty}\,
\left[\psi\left(\frac{2\pi n_2}{k}\right)\,
\sum_{n_1=-\infty}^{+\infty}\delta\left(A_1-\frac{2\pi n_2}{k}-2\pi n_1\right)
\right]=\psi(A_1)\ \ \ .
\end{equation}

However, since this equation possesses $k$ distinct linearly independent
solutions given by
\begin{equation}
<A_1|r>=\psi_r(A_1)=\frac{2\pi}{k}\,C_r\,\sum_{n=-\infty}^{+\infty}\,
\delta\left(A_1-\frac{2\pi r}{k}-2\pi n\right)\ \ \ ,\ \ \ 
r=0,1,2,\dots,k-1\ \ \ ,
\end{equation}
where $C_r$ is some normalisation factor, it is clear that there are exactly
$k$ distinct physical states $|r>$ for the quantised U(1) Chern-Simons theory
whose action is normalised with the factor $N_k=\hbar k/(4\pi)$.
Moreover, the normalisation $C_r$ for each of these configuration space
wave functions of physical states is obtained from the obvious
condition,
\begin{equation}
<A_1|\sum_{k_1,k_2=-\infty}^{+\infty}\hat{U}(k_1,k_2)|A_1'>=
\sum_{r=0}^{k-1}\,<A_1|r><r|A_1'>\ \ \ .
\end{equation}
The explicit resolution of this last constraint then finally provides
the following configuration space wave functions for the $k$ physical states
$|r>$ ($r=0,1,2,\dots,k-1$),
\begin{equation}
<A_1|r>=\sqrt{\frac{2\pi}{k}}\,\sum_{n=-\infty}^{+\infty}\,
\delta\left(A_1-\frac{2\pi r}{k}-2\pi n\right)\ \ \ .
\end{equation}

The coherent state wave functions $<r|z>$ of the same states are then
easily obtained, using the associated change of basis specified by
the quantities $<A_1|z>$, namely
\begin{equation}
<r|z>=\int_{-\infty}^{+\infty}dA_1\,<r|A_1><A_1|z>\ \ \ .
\end{equation}
An explicit calculation then finds
\begin{equation}
<r|z>=\left(\frac{2\tau_2}{k}\right)^{1/4}\,e^{-\frac{1}{2}|z|^2}\,
e^{\frac{1}{2}(-iz)^2}\,
\Theta\left[\begin{array}{c}
	r/k \\ 0
	\end{array}\right]
\Big(-ik\sqrt{\frac{\tau_2}{\pi k}}\,z\Big|k\tau\Big)\ \ \ ,
\end{equation}
where the $\Theta$-function with characteristics is defined by\cite{Alvarez}
\begin{equation}
\Theta\left[\begin{array}{c}
	\alpha \\ \beta
	\end{array}\right]
\Big(z\Big|\tau\Big)=\sum_{n=-\infty}^{+\infty}\,
e^{i\pi\tau(n+\alpha)^2+2i\pi(n+\alpha)(z+\beta)}\ \ \ .
\end{equation}

With respect to modular transformations, two useful identities for these
$\Theta$-functions are,
\begin{equation}
\Theta\left[\begin{array}{c}
	r/k \\ 0
	\end{array}\right]
\Big(x\Big|k(\tau+1)\Big)=e^{i\pi r^2/k}\,
\Theta\left[\begin{array}{c}
	r/k \\ 0
	\end{array}\right]
\Big(x\Big|k\tau\Big)\ \ \ ,\ \ \ {\rm only\ if}\ k\ {\rm is\ even}\ \ \ ,
\end{equation}
\begin{equation}
\Theta\left[\begin{array}{c}
	r/k \\ 0
	\end{array}\right]
\Big(x\Big|-\frac{k}{\tau}\Big)=
\frac{1}{k}\,\left(-ik\tau\right)^{1/2}\,e^{i\pi\tau r^2/k}\,
\sum_{r'=0}^{k-1}\,e^{2i\pi rr'/k}\,
\Theta\left[\begin{array}{c}
	r'/k \\ 0
	\end{array}\right]
\Big(-\tau x\Big|k\tau\Big)\ \ \ .
\end{equation}

\end{document}